\documentclass[12pt]{article}
\usepackage{cite,hyperref}
\usepackage{amssymb}

\begin{document}
\makeatletter
\@addtoreset{equation}{section}
\renewcommand{\theequation}{\thesection.\arabic{equation}}
\makeatother

\title{Finite-Size Correlation Length and Violations of Finite-Size 
Scaling
}

\author{
  \\
  { Sergio Caracciolo and Andrea Gambassi }              \\
  {\small\it Scuola Normale Superiore and INFN -- Sezione di Pisa}  \\[-0.2cm]
  {\small\it I-56100 Pisa, ITALIA}          \\[-0.2cm]
  {\small Internet: {\tt sergio.caracciolo@sns.it}, {\tt
  andrea.gambassi@sns.it}  }    
  \\[-0.1cm]  \and
  { Massimiliano Gubinelli}              \\
  {\small\it Dipartimento di Fisica and INFN -- Sezione di Pisa}    \\[-0.2cm]
  {\small\it Universit\`a degli Studi di Pisa}        \\[-0.2cm]
  {\small\it I-56100 Pisa, ITALIA}          \\[-0.2cm]
  {\small Internet: {\tt mgubi@cibs.sns.it}}   
  \\[-0.1cm]  \and
  { Andrea Pelissetto}              \\
  {\small\it Dipartimento di Fisica and INFN -- Sezione di Roma I}
  \\[-0.2cm] 
  {\small\it Universit\`a degli Studi di Roma ``La Sapienza"}        \\[-0.2cm]
  {\small\it I-00185 Roma, ITALIA}          \\[-0.2cm]
  {\small Internet: {\tt Andrea.Pelissetto@roma1.infn.it}}   \\[-0.2cm]
  {\protect\makebox[5in]{\quad}}  
  \\
}

\newcommand{\be}{\begin{equation}}
\newcommand{\ee}{\end{equation}}
\newcommand{\<}{\langle}
\renewcommand{\>}{\rangle}
\newcommand{\R}[2]{$(#2,#1)$}
\def\spose#1{\hbox to 0pt{#1\hss}}
\def\ltapprox{\mathrel{\spose{\lower 3pt\hbox{$\mathchar"218$}}
 \raise 2.0pt\hbox{$\mathchar"13C$}}}
\def\gtapprox{\mathrel{\spose{\lower 3pt\hbox{$\mathchar"218$}}
 \raise 2.0pt\hbox{$\mathchar"13E$}}}
\newcommand{\reff}[1]{(\ref{#1})}

\newcommand{\co}{ {\cal O}}
\newcommand{\im}[1]{\left\lfloor #1 \right\rfloor}
\newcommand{\hq}{\hat{q}}
\newcommand{\ve}[1]{{\mathbf{#1}}}
\def\bsigma{\mbox{\protect\boldmath $\sigma$}} 
\def\hatp{\hat p}
\def\hatq{\hat q}

\maketitle
\thispagestyle{empty}   


\begin{abstract}
We address the problem of the definition of the 
finite-volume correlation length.
First, we study the large-$N$ limit of the $N$-vector model, and we show
the existence of several constraints on the definition if  
regularity of the finite-size scaling functions and  
correct anomalous behaviour above the upper critical dimension are required. 
Then, we study in  detail a model in which the zero mode in
prohibited. Such a model is a generalization of the fixed-magnetization 
Ising model which is equivalent to the lattice gas. 
Also in this case, we find that the finite-volume correlation length
must satisfy appropriate constraints in order to obtain regular 
finite-size scaling 
functions, and, above the upper critical dimension, an
anomalous scaling behaviour. 
The large-$N$ results are confirmed by a one-loop calculation in the lattice 
$\phi^4$ theory.
\end{abstract}

\clearpage

\section{Introduction}

Phase transitions are characterized by a nonanalytic behaviour at the 
critical point. These non-analyticities are however observed only in the 
infinite-volume limit. If the system is finite, all thermodynamic functions
are analytic in the thermodynamic parameters, temperature, applied magnetic 
field, and so on. However, even from a finite sample, it is possible to 
obtain many informations on the critical behaviour.
Indeed, large but finite systems show a universal behaviour called 
finite-size scaling (FSS). The FSS hypothesis, formulated for the
first time by Fisher~\cite{Fisher71,Fisher72,Imry} and justified 
theoretically by using renormalized continuum field theory 
\cite{Brezin82,Brezin85} (a collection of
relevant articles on the subject appears in~\cite{Cardy88}), is a very
powerful method to extrapolate  to the thermodynamic limit
the results obtained from a finite sample, both in experiments and in 
numerical simulations. In particular, the most recent Monte Carlo 
studies rely heavily on FSS for the determination of critical 
properties (see, e.g., 
\cite{LWW-91,Hasenbusch-93,BK-93,BLH-95,Caracciolo95,Caracciolo95b,%
Salas:1997bh,Ferreira:1998tq,HPV-99,BFMMPR-99} 
for recent applications in two and three dimensions; 
the list is of course far from being exhaustive).

The most commonly studied systems are those that show a second-order 
phase transition at a critical temperature $T_c$.
In this case there are quantities $\co$ which in infinite volume behave as 
\begin{equation}
  \co_\infty(t)\equiv \langle \co \rangle_\infty  \sim |t|^{-x_\co} 
\qquad \hbox{ for }\quad t \to 0
 \label{coinf}
\end{equation}
where $t \equiv (T - T_c)/T_c$ is the reduced temperature and $\langle
\cdot \rangle_\infty$ denote averaging in the infinite-volume system.
In a finite sample of length $L$ in all directions, the 
finite-volume mean values 
$\co_L(t) \equiv \langle \co \rangle_L$ are analytic functions of $t$. 
However, for large $L$, the FSS theory predicts a scaling behaviour of the form
\begin{equation}
 \co_L(t) = G_\co(t, L) \approx L^{x_\co / \nu}\, g_\co \left (t
   L^{1/ \nu} \right). 
\label{co}
\end{equation}
where $\nu \equiv x_\xi$ is the critical exponent which, according
to~(\ref{co}), controls the divergence of the bulk correlation length $\xi$.
The function $g_\co(z)$ is finite and non-vanishing in zero, and should
satisfy
\begin{equation}
   g_\co(z) \sim  |z|^{-x_\co}  \qquad \hbox{ for }\quad z \to \infty
\end{equation}
in order to recover the singular behaviour~\reff{coinf} in the limit
$L\rightarrow\infty$ followed by $t\to 0$.

Equation \reff{co} can be conveniently rewritten in terms 
of the bulk correlation
length,  which avoids the knowledge of the critical temperature:
\begin{equation}
  \co_L(t)  \approx L^{x_\co / \nu}\, f_\co \left({\xi_\infty(t) \over L}
   \right), \label{forma}
\end{equation}
where $ f_\co(z)$ is finite for $z\to\infty$ and 
\begin{equation}
   f_\co(z) \sim  |z|^{x_\co/\nu}  \qquad \hbox{ for }\quad z \to 0.
\end{equation}
Eq.~\reff{forma} shows that, even if two lengths,
the size of the sample and the correlation length,
characterize the finite-sample behaviour
of the observable $\co$, 
in the FSS regime only their ratio is relevant.

Eqs.~(\ref{co}) and~(\ref{forma}) apply only below the upper critical 
dimension $d_>$. 
Indeed, Fisher \cite{Fisher-Temple,Fisher-83,Privman-Fisher-83} 
showed that for $d \ge d_>$
there are violations to FSS. In the context of the renormalization
group these violations can be understood as due to the presence of
dangerously irrelevant operators.
The failure of standard FSS can also
be understood phenomenologically.
For instance, let us consider the order parameter
$M$ for an Ising model in a finite system of size $L$ with 
periodic boundary conditions. 
According to the previous discussion, we expect that 
\be
\< |M| \>_L \sim  L^{- {\beta / \nu}}\, g_M
  \left(t L^{1/\nu}
   \right), \label{gm}
\ee 
while the ordering susceptibility 
\be
 \chi_L = {1\over T}\, L^d \,  \< M; M \>_L \label{chiL}
\ee
behaves as
\be
 \chi_L  \sim  L^{\gamma / \nu}\, g_\chi
  \left( t L^{1/\nu}   \right).  \label{chi}
\ee
Note that, since $\<M\>_L = 0$, we have $\<M^2\>_L = T L^{-d}\chi_L$. 
Moreover, in the low-temperature phase, for $t\to0^-$ at fixed large $L$,
$\< |M| \>_L^2\sim \<M^2\>_L$. Then, if $g_M(0)$ and $g_\chi(0)$ are finite,
relations \reff{gm} and \reff{chi} are compatible only
if the hyperscaling relation~\cite{Fisher74}
\be
             d\, \nu = \gamma + 2\,\beta \label{hs}
\ee
holds. But for $d\ge d_>=4$, hyperscaling is violated, since 
the system shows mean-field behaviour.
The anomalous scaling behaviour can be obtained by adding power-law 
violations. Following~\cite{Young,Leung92} we write
\begin{eqnarray}
\chi_L &=& L^{2+y_2} \tilde{g}_\chi \left(t\,L^{2+y_1}\right) \, ,
\\
\<|M|\>_L & = & L^{-1+y_3} \tilde{g}_{M} \left(t\,L^{2+y_1}\right)\, ,
\label{y3} \\
\xi_L &=& L^{1+y_4} \tilde{g}_\xi \left(t\,L^{2+y_1}\right)\, .
\label{y4}  
\end{eqnarray}
Requiring mean-field behaviour for $L\to\infty$, i.e. 
$\chi_\infty\sim |t|^{-1}$, $\<|M|\>_\infty\sim (-t)^{1/2}$ for $t<0$, 
and $\xi_\infty\sim |t|^{-1/2}$ we obtain 
$y_2=  y_1$, $y_3=-y_1/2$, and $y_4=y_1/2$. 
Using as before $\chi_L = T^{-1} L^d \<M^2\>_L \sim L^d\<|M|\>_L^2$ for 
$t\to 0^-$, we 
obtain a relation which replaces the hyperscaling one~\reff{hs}:
\be -2 - y_1 = 2 + y_1 -d, \ee
that is
\be
y_1 = {d-4\over 2}.
\ee
For the correlation length, this implies 
\be
\xi_L \sim L^{d/4}\,  \tilde{g}_\xi(t L^{d/2})\, .
\label{anomalous}
\ee
Note that the scaling functions now depend  on the combination 
$t L^{d/2}$, which can be interpreted~\cite{Young} as the appearance
of a new {\em thermodynamical}
scale that behaves as $t^{-\tilde{\nu}}$, $\tilde{\nu} = 2/d$.

Anomalous scaling can also be understood in the framework of 
renormalized continuum field theory. In general, in a finite volume,
the allowed momenta are quantized---for instance 
$\ve{p} \in \left( 2 \pi {\mathbb Z}/L \right)^d$ on a hypercubic 
lattice---and therefore,
all nonzero momenta
have an effective mass of order $L^{-1}$ even at the critical point.
Therefore, they can be treated perturbatively. The zero mode must instead be
considered exactly~\cite{Brezin85,Rudnick}. For $d>d_>$,
anomalous FSS can be obtained by taking into account only
the zero mode and completely neglecting the modes with $\ve{p}\not=0$.
However, this  cannot always be done.
Indeed, if an observable is not a zero-momentum quantity, 
nonzero modes must be properly treated. 
This can be the case of the correlation length. 

A question we will ask is whether the finite-volume 
correlation length 
scales anomalously according to \reff{anomalous} as predicted by both
field-theoretical approach  and  phenomenological arguments. 
Indeed the anomalous scaling depends crucially on the finite-volume 
definition one is choosing. For generic choices of the correlation length,
one does not observe a scaling like \reff{anomalous}: specific properties
must be satisfied in order to observe anomalous scaling. 

The question of the FSS behaviour of quantities defined for 
nonzero values of the momenta becomes even more important 
when one considers systems in which the zero mode is not allowed 
because of conservation laws. We are thinking here of the Ising
model at fixed zero magnetization or of its equivalent counterpart, 
the lattice gas with fixed number of particles at half filling. 
The field theory of this model was studied
by Eisenriegler and Tomaschiz~\cite{Eisenriegler}
for $d<4$. They show that one can perform an $\epsilon$-expansion
with $\epsilon = 4-d$ and that the FSS functions of these models
are different from those of the models in which the zero mode is present.
Again we ask whether these field-theoretical results apply to all 
possible definitions of finite-volume correlation length.
As in the previous case, we find that specific properties must be satisfied in 
order to have regular FSS functions and to observe anomalous scaling behaviour
above the upper critical dimension. Note that the conditions 
necessary in this class of models without zero mode are different from 
those in the models without conservation laws. In particular 
``good" definitions in the unconstrained model are ``bad" definitions 
in the model at fixed magnetization.

Our conclusions are supported by a large-$N$ calculation in the 
$N$-vector model and by an explicit one-loop computation in the lattice 
$\phi^4$ model.

\section{$N$-vector model in the large-$N$ limit}

In order to have a complete control of the critical behaviour in the
non-perturbative regime, we shall consider the large $N$-limit of
the $N$-vector $\sigma$-model,  already studied by
Br\'ezin~\cite{Brezin82} (for a detailed analysis at the
lower critical dimension $d=d_<=2$, see~\cite{Caracciolo98}).

We consider a $d$-dimensional hypercubic lattice 
of finite extent $L$ in all directions, unit $N$-vector 
spins $\bsigma$ defined at the sites of the lattice interacting with 
Hamiltonian
\be
{\cal H} \equiv\, - N \sum_{x,y} J(x-y) \bsigma_x \cdot \bsigma_y  - 
               N h \sum_{x} \sigma_x^1,
\label{eq2.1}
\ee
where $J(x)$ is a short-range coupling. The partition function is simply
\be
Z \equiv\, \int \prod_x [d\bsigma_x \, \delta(\bsigma_x^2-1)] \;
e^{-{\cal H}/T}. 
\ee
In the large-$N$ limit, 
assuming periodic boundary conditions and $h=0$, 
the theory is solved in terms of the gap equations
\begin{eqnarray}
\lambda_L \sigma_L = 0, \qquad\qquad
{1\over T} = {\sigma_L^2\over T} + 
   {1\over L^d} \sum_{q\in \Lambda^*_L} 
  {1\over \hq^2 + \lambda_L },
\label{infty}
\end{eqnarray}
where $\sigma_L$ and $\lambda_L$ are the constant values, respectively,
  of  the field $\bsigma$, and of the Lagrange multiplier which
  implements the constraint on the length of the spin, at the
  saddle-point. Moreover, $\hq^2 = - 2 (\widehat{J}(q) - \widehat{J}(0))$, 
$\widehat{J}(q)$ is the Fourier transform of $J(x)$, and $\Lambda^*_L$ is the 
lattice 
\be
\Lambda_L^* \equiv \left( {2\pi\over L}\, \mathbb{Z}_{L}\right)^d.
\ee
In infinite volume, the same equations hold, with the simple 
substitution of the summation with the normalized integral over the first 
Brillouin zone $[-\pi,\pi]^d$. 

The meaning of the parameters $\lambda_L$ and $\sigma_L$ is clarified by
considering the magnetization and the two-point function. 
We define 
\be
M_L \equiv\, \< \sigma^1_0 \>_L, \qquad\qquad G_L(x) \equiv \<
\bsigma_0\cdot\bsigma_x\>_L. 
\ee
Then 
\be
M_L = \sigma_L, \qquad\qquad \widehat{G}_L(q) = {T\over \hq^2 + \lambda_L},
\ee
where $\widehat{G}_L(q)$ is the Fourier transform of $G_L(x)$. Note that the 
susceptibility is given by 
\be
\chi_L = {1\over T} \widehat{G}_L(0) = {1\over \lambda_L}.
\ee
We wish now to define the correlation length. In infinite volume there 
are (at least) two possibilities:
\begin{eqnarray}
\xi^{\rm (exp)}_\infty &\equiv& 
  - \lim_{|x|\to\infty} {|x|\over \log G_\infty(x)}, 
\\
\xi^{\rm (2)}_\infty   &\equiv& 
  \left( {1\over 2d} {\sum_x G_\infty(x) |x|^2\over \sum_x G_\infty(x)} 
  \right)^{1/2} =\, 
\left[ {-1\over 2 d T\chi_\infty} 
   \left. 
   {d^2 \widehat{G}_\infty(q)\over dq_\mu dq_\mu}\right|_{q=0}\right]^{1/2}.
\end{eqnarray}
For $T\to T_c$ the ratio of the two definitions become one---this 
is a peculiarity 
of the large-$N$ limit; in general their ratio is a generic
constant---and thus we do not need to distinguish between them. 
We have 
\be
\xi_\infty = {1\over\lambda_\infty^{1/2}}(1 + O(\lambda_\infty)).
\ee
This expression is exact for $\xi^{\rm (2)}_\infty$.

In a finite box there is no {\em a priori} natural definition of 
correlation length. Clearly, the exponential correlation length
cannot be generalized, so that we must devise definitions that 
converge to $\xi^{\rm (2)}_\infty$ as $L\to \infty$. 
For instance, we can define 
\be
\xi^{(2a)}_L \equiv \left[
  {\widehat{G}_L(0) - \widehat{G}_L(q_{\rm min}) \over 
      \hq^2_{\rm min} \widehat{G}_L(q_{\rm min})}\right]^{1/2},
\ee
where $q_{\rm min} = (2\pi/L,0,\ldots,0)$ (of course, by cubic symmetry,
there are $2d$ equivalent definitions of $q_{\rm min}$), 
which converges to $\xi^{(2)}_\infty$
as $L\to \infty$ (this property is not specific of the large-$N$ limit; 
indeed, 
to prove $\xi_L\to \xi_\infty$ we need only the cubic symmetry of the model).
The definition is motivated by the desire to have, also for finite $L$,
the same relation between $\xi$ and $\lambda$. Indeed, we have
\be
\xi^{(2a)}_L = {1\over \lambda_L^{1/2}},
\ee
exactly. However, an equally valid definition is 
\be
\xi^{(2b)}_L \equiv \left[
  {\widehat{G}_L(0) - \widehat{G}_L(q_{\rm min}) \over
      \hq^2_{\rm min} \widehat{G}_L(0)}\right]^{1/2},
\ee
which gives 
\be
\xi^{(2b)}_L = \left(\lambda_L + \hq^2_{\rm min}\right)^{-1/2},
\ee
or 
\be
\xi^{(2c)}_L \equiv \left[
  {\widehat{G}_L(0) - \widehat{G}_L(q_{\rm min}) \over
      \hq^2_{\rm min} 
      (2 \widehat{G}_L(q_{\rm min}) - \widehat{G}_L(0)) }\right]^{1/2},
\ee
which gives
\be
\xi^{(2c)}_L = \left(\lambda_L - \hq^2_{\rm min}\right)^{-1/2}.
\ee
For $L\to \infty$ at $t$ fixed (i.e. $\lambda_L$ fixed)
\be
\xi^{(2a)}_L \approx  \xi^{(2b)}_L \approx \xi^{(2c)}_L \to \xi^{(2)}_\infty,
\ee
so that all of them are correct. But the question is whether all of them 
have the correct FSS properties. 

We will now show that $\xi^{(2c)}_L$ is a bad definition. In the large-$N$
limit, this is evident. Indeed, even for finite $L$, $\xi^{(2c)}_L$ 
becomes infinite for $\lambda_L = \hq^2_{\rm min}$. This implies 
that the FSS scaling function $g_\xi(z)$ defined in \reff{co} diverges 
at a finite value of $z$ and is not defined below it. Note that this is not 
a peculiarity of the large-$N$ limit, but is completely general and it is due 
the fact that $\xi^{(2c)}_L$ is not defined for all temperatures at finite 
$L$. For sufficiently low values of $T$ one always has 
$2 \widehat{G}_L(q_{\rm min}) < \widehat{G}_L(0)$ (indeed, 
for $T=0$ the system orders and only $\widehat{G}_L(0)$ is nonzero), 
so that the square root becomes ill defined.
Thus, the first necessary requirement is to have a definition that 
is valid  for all temperatures.

Now, we should discuss whether $\xi^{(2a)}_L$ and $\xi^{(2b)}_L$,
which satisfy the previous requirement, can both be used as 
finite-size correlation lengths.
Below the upper critical dimension, one can show---this is a trivial 
generalization of the results of \cite{Brezin82}---that indeed
both of them have correct FSS properties. 
However, above the upper critical dimension, i.e. for $d>4$, this is
no longer true.  
We will now show it explicitly in the large-$N$ limit, by reviewing 
critically the results by Br\'ezin \cite{Brezin82}.

\subsection{Infinite-volume solution above the upper critical dimension}

The infinite-volume solution is well known. The system has a critical 
point at $T_{c}$
\be
{1\over T_{c}} \equiv \int_{-\pi}^\pi\, {d^dq\over (2 \pi)^d} {1\over
  \hat{q}^2 } =  {\cal C}_{d,0},
\label{Tc}
\ee
where we define, for $d>2n+2$,
\be
 {\cal C}_{d,n} \equiv  \int_{-\pi}^\pi\, {d^dq\over (2 \pi)^d} 
   {1\over \left( \hat{q}^2 \right)^{n+1}}.
\ee
Then, for $t> 0$,
\be
 \xi_\infty \approx \sqrt{{ {\cal C}_{d,1} \over  {\cal C}_{d,0}}\,{1 \over
     t}}, \qquad
 \chi_\infty = \frac{1}{T}\,\widehat{G}_\infty(0) = \xi_\infty^2 \approx 
   { {\cal C}_{d,1} \over {\cal C}_{d,0}}\,{1 \over t} ,
\ee
while, for $t<0$, the magnetization is non-vanishing and 
\be 
M_\infty^2 = 1 - {T\over T_c} = - t.
\ee

\subsection{Finite-size scaling above the upper critical dimension} 
\label{FVWZM}

In finite volume there is no critical point 
and indeed $\lambda_L > 0$ for all $T>0$ and 
$\lambda_L \to 0$ for $T\to 0$. We can rewrite the gap 
equation
\begin{eqnarray}
{1\over T} 
&=& \frac{1}{L^d \lambda_L} + 
 \frac{1}{L^d}\sum_{q\in\Lambda_{L}^* -\{0\}}\frac{1}{\hq^2+\lambda_L} 
 =   \frac{1}{L^d \lambda_L} +   \int_{-\pi}^\pi\, 
{d^dq\over (2 \pi)^d}   {1\over \hat{q}^2 +   \lambda_L } +
 \Psi(\lambda_L,L),
 \label{poi} 
\end{eqnarray}
where the function $\Psi(\lambda_L,L)$ is defined in \reff{defPsi}. Then
\be
\frac{1}{T_c}-\frac{1}{T} =   \int_{-\pi}^\pi\, {d^dq\over (2
 \pi)^d}   \left( {1\over \hat{q}^2} -  {1\over \hat{q}^2 +
 \lambda_L }\right)  -\frac{1}{L^d\lambda_L} 
- \Psi(\lambda_L,L),
\ee
so that, for $\lambda_L\to 0$, $L\to \infty$, $\lambda_L L^2$ finite, we have
\be
t\, {\cal C}_{d,0} \approx  \lambda_L\, {\cal{C}}_{d,1}-  L^{2-d} \left[
\frac{1}{\lambda_L L^2} +  {\cal I}(\lambda_L L^2)\right] ,
  \label{difff} 
\ee
where $\cal I(\rho)$ is defined in \reff{defcalI}.
In App.~\ref{usefulfunc} we show that ${\cal I}(\rho)$ is not singular in zero.

\subsubsection{Trivial scaling of $\xi^{(2a)}$}
\label{TrivialScaling}

Let us consider \reff{difff} and replace $\lambda_L$ with 
$\xi^{(2a)}_L$, which we simply write as $\xi_L$. Then
\be
t\, L^2\, {\cal C}_{d,0}  =  \left({L^2\over \xi_L^2}\right)\, {\cal
  C}_{d,1} -  L^{4-d} \left[ 
\frac{\xi_L^2}{ L^2} + {\cal I}\left(\frac{L^2}{\xi_L^2}\right)\right] .
\label{relation}
\ee
We wish now to consider the standard FSS, i.e. 
the limit $t\to 0$, $L\to \infty$, with $t L^{1/\nu} = t L^2$ fixed.
For $t>0$, \reff{relation} implies $\xi_L \sim L$ and
\be
t\, L^2\, {\cal C}_{d,0}  =  \left({L^2\over \xi_L^2}\right)\, {\cal
  C}_{d,1} + O\left(  L^{4-d} \right).
\label{eq:2.26}
\ee
This can be written as 
\be
\xi_L = L g_\xi (t L^2),
\ee
where
\be
g_\xi(z) = \sqrt{{{\cal C}_{d,1} \over {\cal C}_{d,0}} {1\over z}}.
\label{high}
\ee
However, the $L$-dependence is only apparent, and \reff{eq:2.26} 
simply states the trivial fact $\xi_L = \xi_\infty + O(L^{4-d})$.

For $t<0$, it is not possible that $\xi_L\sim L$. Indeed, the 
l.h.s. of \reff{relation} is negative, while the only negative 
term in the r.h.s. is the second one. Therefore, $\xi_L^2 \sim L^{d-2}$ 
---and thus $\xi_L/L\to\infty$ since $d>4$ ---and 
\be
t\, L^2\, {\cal C}_{d,0}  = -  \left({ \xi_L^2\over L^{d-2}}\right) + 
O\left(  L^{4-d} \right).
\ee
Therefore
\be
\xi_L = L^{\frac{d-2}{2}} \sqrt{- t L^2 {\cal C}_{d,0}}
\label{eqascal}
\ee
which is the correct behaviour of the finite-size correlation length
in the broken phase because of the Goldstone mode, but it is
not in agreement with the usual FSS behaviour.

Similar results hold for the magnetization in the low-temperature phase, since
\be
\<|M|\>^2_L \sim L^{-d} \chi_L = L^{-d} \xi_L^2 \sim L^{-2} \left(-t L^2
\right) =  -t = M^2_\infty,
\ee
which means that the FSS for this quantity is trivial.

\subsubsection{Anomalous finite-size scaling of $\xi_L^{(2a)}$}
\label{AnomalousScaling} 

In the previous Section, we have shown that it is possible
to define mathematically---although it is physically irrelevant---standard FSS,
even for $d>4$. However, such a scaling is possible only for 
$t>0$, and, correspondingly, the FSS functions are singular for 
$t L^2\to 0$. 

Now, we wish to obtain a different FSS that is defined for all values of the 
temperature. As we shall see, this is the standard anomalous scaling.
Going back to \reff{relation}, in order to have relations that are 
valid for all $t$, we must consider a limit in which both the 
first and the second term in the right-hand side of~\reff{relation} are 
present. Therefore, we must require
\be
t L^2 \sim L^2/\xi_L^2 \sim L^{2-d} \xi_L^2,
\ee
which implies $\xi_L\sim L^{d/4}$, $t \sim L^{-d/2}$. Then, we rewrite 
\reff{relation} in the limit $t\to 0$, $L,\xi_L\to \infty$ at fixed 
$t L^{d/2}$ and $\xi_L L^{-d/4}$. We obtain
\be
t L^{d/2} {\cal C}_{d,0} = {L^{d/2}\over \xi_L^2} {\cal C}_{d,1} - 
   {\xi_L^2\over L^{d/2}} + \, O(L^{(4-d)/2}).
\ee
Solving this equation we obtain
\be
\xi_L = L^{d/4} \tilde{g}_\xi\left(t L^{d/2} \right),
\label{eqxiinfty}
\ee
where
\be
 \tilde{g}_\xi(z) = 
 \left[ - {{\cal C}_{d,0} z\over 2} + {1\over 2} 
    \sqrt{{\cal C}_{d,0}^2 z^2 + 4 {\cal C}_{d,1}} \right]^{1/2},
\label{eq2.38}
\ee
consistently with \reff{anomalous}. Note that $\tilde{g}_\xi(z)$ 
is regular in zero, and that for $z\to+\infty$ one recovers the behaviour
\reff{high},  and for $z\to-\infty$ the behaviour~\reff{eqascal}.

\subsubsection{Finite-size scaling of $\xi^{(2b)}_L$} 

Let us now consider the second definition of correlation length. For
$L\to \infty$ 
\be
\xi^{(2b)}_L\approx \left(\lambda_L + {4\pi^2\over L^2}\right)^{-1/2}.
\label{eq:2.44}
\ee
Since, for $\lambda_L\ge 0$,
\be
\left(\lambda_L + {4\pi^2\over L^2}\right)^{-1/2} \le {L\over 2\pi},
\ee
it is clear that $\xi^{(2b)}_L$ cannot scale anomalously, i.e. as
$L^{d/4}$. We are thus forced to consider 
$\xi_L^{(2b)} \sim L$. However, at variance with the previous case, 
this condition does not fix the scaling of $\lambda_L$, and thus,
we can consider both the trivial scaling and the anomalous one. 
Indeed, if we take $t\to 0$, $L\to \infty$, with $t L^2$ fixed,
then $\lambda_L L^2\to t L^2 {\cal C}_{d,0}/{\cal C}_{d,1}$ for $t>0$
and $\lambda_L L^2\to 0$ for $t<0$. Then
\be
\xi_L^{(2b)} = L g_\xi^{(2b)} (t L^2),
\ee
where
\be
g_\xi^{(2b)} (z) = \cases{(4 \pi^2 + z {\cal C}_{d,0}/{\cal C}_{d,1})^{-1/2}
             & \qquad for $z > 0$, \cr
          (2 \pi)^{-1} & \qquad for $z < 0$.}
\ee
Therefore, the standard FSS limit exists, but the FSS function is not smooth
at the origin. Moreover, as we observed in the case of $\xi^{(2a)}_L$, 
it does not give any information on the finite-size physics of the system,
since it is simply obtained by replacing $\lambda_L$ by $\lambda_\infty$ 
in \reff{eq:2.44}, and then using the expression of $\lambda_\infty$ in terms
of $t$.

If instead we keep $t L^{d/2}$ fixed, then $\lambda_L L^2\to 0$, so that 
\be
\xi_L^{(2b)} = {L\over 2\pi}.
\ee
The FSS function is well defined but does not contain any dynamical 
information.

\subsection{Summary}

In this Section we discussed the FSS behaviour of the correlation length 
above the upper critical 
dimension. First, we considered 
the standard scaling $t\to 0$, $L\to \infty$ at fixed $t L^2$. 
The limit exists, but it is singular for $t L^2 = 0$ and trivial,
since the corresponding FSS functions do not contain any dynamical 
information.
A nontrivial result is obtained only by considering the anomalous 
scaling at fixed $t L^{d/2}$ for some specific definitions of 
finite-volume correlation length. Indeed,
for {\em generic} definitions, 
for instance for $\xi^{(2b)}_L$,
however the limiting behaviour of the temperature 
is defined---keeping $t L^2$ or $t L^{d/2}$ fixed---the FSS is always trivial:
the FSS functions do not contain any information on the dynamics of the 
system. In particular, there is no way to observe the anomalous scaling 
$\xi_L\sim L^{d/4}$. Such a scaling can only be observed by an appropriate
choice of $\xi_L$, and indeed, the definition 
$\xi^{(2a)}_L$ does show an anomalous scaling behaviour. 
In the large-$N$ limit
it is easy to understand why: since $(\xi^{(2a)}_L)^2 = \chi_L$, 
$\xi^{(2a)}_L$ is effectively a zero-momentum quantity. 
Of course, the interesting question is whether these results apply to generic 
values of $N$. Since the theory above the critical dimension is 
essentially Gaussian, we expect $\xi^{(2a)}_L$  to show anomalous 
scaling for all values of $N$, and this is 
confirmed by the one-loop computation that will be presented in
Sec.~\ref{sec4}.  
Of course, there are many other definitions
that also show anomalous scaling. If one follows carefully the
computations of Sec.~\ref{AnomalousScaling}, one can convince herself
that the relevant  
property is that $\xi_L$ behaves as $\lambda^{-1/2}_L$ for 
$\lambda_L\to 0$. For instance, we could have used any definition 
of the form $\xi_L = \lambda^{-1/2}_L h(\lambda_L L^\rho)$, 
with any $\rho\ge 0$, $h(\infty)=1$, and $h(0)$ finite, again obtaining 
anomalous scaling. In other words, in order to have an anomalous scaling
behaviour $\xi_L\sim L^{d/4}$ and nontrivial scaling functions, 
the finite-volume correlation length
must be finite for all temperatures $T\not=0$ and diverge properly 
for $T\to 0$.

\section{The model at fixed zero magnetization in the large-$N$ limit}

As we mentioned in the introduction, in order to discuss the FSS of a lattice 
gas, one can consider the Ising model at fixed magnetization. 
To observe a critical behaviour one must fix the magnetization to 
zero. Here, we discuss the natural generalization for $N\to \infty$, 
considering in a finite box a model with a modified gap equation of the form
\be
{1\over T} =  \frac{1}{L^d}\sum_{q\in\Lambda_{L}^* -\{0\}}
  {1\over \hat{q}^2 +   \lambda_L}.
\label{FVgapeq}
\ee
The only difference with respect to the case we have considered above is 
the absence of the zero mode in the summation.
In infinite volume we recover the usual gap equation and therefore 
the same results apply. In finite volume,
for $T\ge 0$\ the gap equation has always a solution
$\lambda_L\ge - \hat{q}^2_{\rm min}$.
In particular $\lambda_L\to - \hat{q}^2_{\rm min}$ for 
$T\to 0$.

Since in this model the magnetization is vanishing, we cannot define 
quantities at zero momentum, and therefore we cannot use 
$\xi^{(2a)}_L$ or $\xi^{(2b)}_L$ as our definition of correlation length.
However, even in this case we can define $\xi_L$ in order to obtain 
$\xi_L^2 = 1/\lambda_L$. For instance, we can consider
\be
\xi^{(2d)}_L \equiv \left[
   {\widehat{G}_L(q_2) - \widehat{G}_L(q_1)\over 
    \hq^2_1 \widehat{G}_L(q_1) - \hq_2^2 \widehat{G}_L(q_2)}\right]^{1/2},
\label{xi2d-def}
\ee
where $q_1$ and $q_2$ are two arbitrary non-vanishing momenta in
$\Lambda^*_L$.  
This is the natural generalization of $\xi_L^{(2a)}$, which does not make 
use of the propagator at zero momentum. In the theory with zero mode,
this definition is of course a good one and $\xi^{(2d)}_L$ has all the 
expected F'S properties. However, in the absence of the zero mode, 
this definition has a serious shortcoming.
It is defined only for $\lambda_L> 0$, and we know that $\lambda_L$ 
may become negative for $L$ finite. As a consequence, even in finite volume,
$\xi^{(2d)}_L$ diverges at a pseudo-critical temperature
\be
{1\over T_c(L)} \equiv {1\over L^d} \sum_{q\in \Lambda^*_L-\{0\}}
{1\over \hq^2}. 
\label{TcL}
\ee
In order to overcome these problems we wish to adopt a new definition
such that $\xi_L$ is finite for all temperatures. This means that 
we need a definition that applies for all 
$\lambda_L\ge -\hq_{\rm min}^2$.
Such a property is enough to guarantee the regularity of the FSS 
functions below the upper critical dimension. However, as we discussed
in the previous Section, in order to obtain anomalous scaling
above the upper critical dimension, the finite-volume correlation length
must diverge for $T\to 0$. Thus, we require that 
for $\lambda_L\to - \hq^2_{\rm min}$ 
\be
\xi_L \to (\lambda_L + \hq^2_{\rm min})^{-1/2}.
\label{fixedM-scalingxi}
\ee
It is easy to write down a definition that satisfies \reff{fixedM-scalingxi}
exactly. For instance, if we define 
\be
\xi_L^{(2e)} \equiv \left[{\widehat{G}_L(q_{\rm min}) - \widehat{G}_L(q_2) \over 
             \widehat{G}_L(q_2) (\hq^2_2 - \hq^2_{\rm min})}\right]^{1/2},
\ee
for any $q_2\not=0,q_{\rm min}\in \Lambda_L^*$, we have exactly 
\be
\xi_L^{(2e)} = (\lambda_L + \hq^2_{\rm min})^{-1/2}.
\ee
We will now discuss the FSS properties of $\xi^{(2d)}_L$ and of 
$\xi_L^{(2e)}$.

\subsection{Finite-size scaling above the upper critical dimension}

Considering for simplicity only the case $4<d<6$, 
we can rewrite the gap equation as
\be
{1\over T} =   {\cal{C}}_{d,0} - \lambda_L {\cal{C}}_{d,1} +
  \widehat{\Psi}(\lambda_L,L),
\ee
where $\widehat{\Psi}(\lambda_L,L)$ is defined in \reff{Psi_N}.
Then, for $\lambda_L\to 0$, $L\to \infty$, $\lambda_L L^2$ finite,
\be
t\, {\cal C}_{d,0} =  \lambda_L\, {\cal{C}}_{d,1} - L^{2-d}  
  \widehat{\cal I}(\lambda_L L^2)  ,
\label{eqdtra4e6}
\ee
where $\widehat{\cal  I}(\rho)$ is defined in~\reff{IN}. 

\subsubsection{Finite-size scaling of $\xi^{(2d)}_L$}

Using $\xi_L=\xi^{(2d)}_L=\lambda_L^{-1/2}$, we rewrite \reff{eqdtra4e6} as 
\be
t\,L^2 {\cal{C}}_{d,0}= \left( \frac{L^2}{\xi_L^2}\right)
{\cal{C}}_{d,1} -L^{4-d} \widehat{\cal I}\left(\frac{L^2}{\xi_L^2}\right), 
\label{eqgapconzeta}
\ee
which should be compared with \reff{relation}.

Keeping $tL^2$\ as the scaling variable and taking the FSS limit we
obtain, for $t>0$, the same results as in
Section~\ref{TrivialScaling}, but  for $t<0$\ there is no scaling form 
since $\xi_L$ is not defined for $t < 0$.

We can also obtain an anomalous scaling. If $L/\xi_L\to 0$, we can rewrite 
\reff{eqgapconzeta} as 
\be
t\,L^{d-2} {\cal{C}}_{d,0}= L^{d-4}\;\frac{L^2}{\xi_L^2}\,
{\cal{C}}_{d,1} - {\cal{I}}(0) + O\left( \frac{L^2}{\xi_L^2}\right) 
\label{scalanomalo}
\ee
where we have used $\widehat{\cal{I}}(0)= {\cal{I}}(0)$, see~\reff{dk}. 
It follows, for $t\,L^{d-2}> -{\cal I}(0) /{\cal C}_{d,0}$,
\be
\xi_L = L^{\frac{d-2}{2}}\,\tilde{g}_\xi^0 (t\,L^{d-2})
\label{eq:3.11}
\ee
with 
\be
\tilde{g}_\xi^0(z) =
 \left[\frac{{\cal{C}}_{d,1}}{{\cal{C}}_{d,0}z + {\cal{I}}(0)}\right]^{1/2},
\ee
which shows a different violation of FSS if compared with~\reff{anomalous}.
However, there is no much physics in this ``new" anomalous scaling. Indeed,
if we define 
\be
\hat{t} = {T - T_c(L) \over T} \approx t + 
    L^{2-d} {{\cal I}(0)\over {\cal C}_{d,0}},
\ee
where $T_c(L)$ is the pseudo-critical temperature \reff{TcL}, then 
\reff{eq:3.11} can be rewritten as 
\be
\xi_L = L^{\frac{d-2}{2}}\, g_\xi(\hat{t} \,L^{d-2}) = 
        L g_\xi(\hat{t} \, L^2) \sim \hat{t}^{-1/2},
\ee
where $g_\xi(z)$ is defined in \reff{high}. The new scaling is just 
a consequence of the definition of the correlation length, but 
gives no information on the finite-volume physics.




\subsubsection{Finite-size scaling of $\xi_L^{(2e)}$}

The correlation length $\xi_L^{(2e)}$ is defined for all values of $\lambda_L$,
i.e. for $\lambda_L > - \hq_{\rm min}^2 \approx -4 \pi^2/L^2$. 
When $\lambda_L L^2 \to - 4 \pi^2$, the function 
$\widehat{\cal I}(\lambda_L L^2)$
becomes singular, see \reff{defL}. It is thus convenient to introduce a 
new function $\widetilde{\cal I}(\lambda_L L^2)$ which is finite in the limit
$\lambda_L L^2 \to - 4 \pi^2$ by writing
\be
\widehat{\cal I}(\rho) = \,2\,d\, \left[\frac{1}{4\pi^2+\rho}+ \widetilde{\cal
    I}(4\pi^2+\rho)  \right].
\label{widetildeI-def}
\ee
Then, for $4<d<6$, we can rewrite \reff{eqdtra4e6}\ as
\be
t\,L^2{\cal{C}}_{d,0} = \left[\left( \frac{L}{\xi_L}\right)^2 - 4\pi^2 \right]
{\cal{C}}_{d,1} - 2d\, L^{4-d}\left[ \left(\frac{\xi_L}{L} \right)^2 +
  \widetilde{\cal I}\left( \frac{L^2}{\xi_L^2}\right)\right] ,
\label{eq3}
\ee
where, in this section, $\xi_L = \xi_L^{(2e)}$.
If we introduce
\be
\tilde{t} \equiv t +\frac{4\pi^2}{L^2}\, \frac{{\cal{C}}_{d,1}}{{\cal{C}}_{d,0}} ,
\ee
we may write \reff{eq3} as 
\be
 \tilde{t}\,L^2{\cal{C}}_{d,0}=\left(\frac{L^2}{\xi_L^2}\right)\,{\cal{C 
}}_{d,1} - 2d\,L^{4-d}\left[\left(\frac{\xi_L}{L}\right)^2 +
\widetilde{\cal I}\left( \frac{L^2}{\xi_L^2}\right)\right]  ,
\label{relationII}
\ee
which is identical to~\reff{relation} with the trivial substitutions 
${\cal{C}}_{d,j}  \to  { {\cal{C}}_{d,j} / 2d }$, $j=0,1$,  and 
${\cal I} \to \widetilde{\cal I}$.
Thus, all considerations we made for 
\reff{relation} apply here. In particular, we can consider the 
FSS limit with $\xi_L L^{-d/4}$ fixed, obtaining
\be
\xi_L = \, L^{d/4} \tilde{g}_\xi (\tilde{t} L^{d/2}),
\label{eq3.anomalo}
\ee
where $\tilde{g}_\xi(z)$ is defined in \reff{eq2.38}. Although formally
the same, \reff{eq3.anomalo} differs significantly from 
\reff{eq2.38} because of the presence of $\tilde{t}$. This means that the 
anomalous scaling is not obtained by varying $t$ as $L^{-d/2}$, but 
rather as
\be
t = - 4 \pi^2 { {\cal C}_{d,1}\over {\cal C}_{d,0}} L^{-2} + 
     {{\rm const}} \, L^{-d/2}.
\ee
The presence of the $L^{-2}$ term represents a sort of finite-volume 
critical-temperature renormalization and, as we will show in the next section, 
is a general phenomenon and not a peculiarity of the large-$N$ limit.

As before, we can obtain the trivial scaling from \reff{eq3.anomalo} 
by taking the limit $\tilde{t} L^{d/2}\to\infty$. 
For example, for $\tilde{t}\to+\infty$, we obtain
\be
\xi_L = L {g}_\xi (\tilde{t} L^2) = L {g}_\xi \left(t L^2
  + 4\pi^2\,   \frac{{\cal{C}}_{d,1}}{{\cal{C}}_{d,0}}\right) ,
\label{scalform3}
\ee
where $g_\xi(z)$ is defined in \reff{high}.

\subsection{Summary}

We have discussed the FSS of the correlation length in a model that 
generalizes for $N\to\infty$ the lattice gas. As before, we find 
several constraints on the finite-volume definition of 
correlation length. First, $\xi_L$ must be defined for all 
temperatures. This condition is of course obvious. What is not obvious 
is that definitions that are ``good" for the model with the zero 
mode are ``bad" here. An example is $\xi_L^{(2d)}$. 
The reason is very general and it is due to the fact that typical 
configurations for $T\to 0$ are different in the two cases. 
For instance, consider $\xi_L^{(2d)}$ with $q_1 = q_{\rm min}$. 
Since, for $T=0$,  $\widehat{G}_L(q)$ 
vanishes except for $q=q_{\rm min}$ or for the $(2d-1)$ equivalent
points, the argument of the square root in \reff{xi2d-def} is negative  
 and therefore  $\xi_L^{(2d)}$ is not defined for low temperatures.

A correlation length that is well defined for all temperatures 
shows standard FSS behaviour below the upper critical dimension. 
However, this is not enough to have an anomalous scaling behaviour. 
Indeed, consider $\xi_L^{(2e)}$ and replace $q_{\rm min}$ by any other 
non-vanishing momentum. Then, it is easy to verify that, 
for $\lambda_L > -\hq_{\rm min}^2$, $\xi_L$ is bounded by $L$, so that 
no anomalous scaling can be observed. As before the important property is 
that $\xi_L$ diverges for $T\to 0$. 

\section{The correlation length in perturbation theory} \label{sec4}

\newcommand{\res}{\mbox{\cal{Res}}}
\newcommand{\bgreek}[1]{\mbox{\boldmath$#1$}}
\newcommand{\bphi}{\mbox{\boldmath$\phi$}}
\newcommand{\dr}{\delta r}
%
We wish now to show briefly that the results we have obtained non-perturbatively
in the large-$N$ limit can be recovered in perturbation theory, 
by considering the $\phi^4$ theory. 

Specifically, we consider $N$-component vector fields on a 
$d$-dimensional hypercubic lattice and the Hamiltonian
\be
H[\bphi] \equiv \sum_{x} \left[ 
 - \frac{1}{2}  \bphi_x\cdot\partial^2 \bphi_x + 
  \frac{1}{2}r \bphi_x\cdot\bphi_x +\frac{u}{4!}(\bphi_x\cdot\bphi_x)^2\right]
\ee
where $\partial^2$ is a lattice Laplacian, such that $\partial^2\to -\hq^2$ 
after a Fourier transform.

At one loop we have, for a finite lattice of size $L$,
\be
\widehat{G}^{-1}_L(q)=\Gamma^{(2)}_L(q) = 
\hat{q}^2 + r + u \Sigma_L(r) + O(u^2),
\label{oneloop}
\ee
where 
\be
\Sigma_L(r) \equiv 
\frac{N+2}{6} \frac{1}{L^d} \sum_{p\in\Lambda^*_L} \frac{1}{\hat{p}^2 + r}.
\ee
In the fixed-magnetization ensemble, we must simply eliminate the 
zero mode. Thus, \reff{oneloop} remains valid with the simple 
substitution of $\Sigma_L(r)$ with $\Sigma_L'(r)$ in which the 
summation over $p$ is restricted to $(\Lambda^*_L - \{0\})$.
We will now discuss the FSS behaviour in perturbation theory of our 
``good" definitions of finite-volume correlation length in the two cases,
above the upper critical dimension.

\subsection{The model with the zero mode}

Using \reff{oneloop}, we have for $\xi_L^{(2a)}$:
\be
\left(\xi^{(2a)}_L\right)^{-2} = r + u \Sigma_L(r) + O(u^2).
\label{GBS4}
\ee
The critical parameter $r_c$\ is defined, 
in the infinite volume limit, by
\be
        r_c + u\Sigma_\infty(r_c) = O(u^2),
\ee
so that $r_c = -u\Sigma_\infty(0) + O(u^2)$.
Then, \reff{GBS4}\ may be written as
\begin{eqnarray}
\left(\xi^{(2a)}_L\right)^{-2} &=& 
r-r_c + u (\Sigma_L(r)-\Sigma_\infty(r_c)) + O(u^2) =
\nonumber\\ 
                    &=& \dr + 
u [\Sigma_L(\dr)-\Sigma_\infty(\dr) + 
   \Sigma_\infty(\dr) - \Sigma_\infty(0)] + O(u^2),
\end{eqnarray}
where we define $\dr \equiv r - r_c$.

Then, for $\delta r\to 0$, $L\to \infty$, with $\delta r L^2$ finite, 
using the results of the Appendix we have
\begin{eqnarray}
\Sigma_L(\dr) - \Sigma_\infty(\dr)&=& 
 \frac{N+2}{6} \left[ \frac{1}{\dr L^d}+ L^{2-d}{\mathcal{I}}(\dr L^2)\right],
\\
\Sigma_\infty(\dr) - \Sigma_\infty(0) &=& 
\frac{N+2}{6}(-\dr)\;{\mathcal{C}_{d,1}}.
\end{eqnarray}
It follows
\be
\left(\xi^{(2a)}_L\right)^{-2} =\, 
\dr \left[1-\frac{N+2}{6}u\;{\mathcal{C}}_{d,1}\right] + 
\frac{N+2}{6}u\;
\left[ \frac{1}{\dr L^d} + L^{2-d}{\mathcal{I}}(\dr L^2)\right] + O(u^2),
\label{eqxi2a_tII}
\ee
i.e.
\be
\xi^{(2a)}_L = L^{d/4}\;\tilde{g}_\xi^\phi(\dr L^{d/2}),
\label{scalingformxi2a}
\ee
where
\be
\left[\tilde{g}_\xi^\phi(z)\right]^{-2} = \,
   z\left[1-\frac{N+2}{6}u\;{\mathcal{C}}_{d,1}\right] +  
             \frac{N+2}{6}u\;\frac{1}{z} + O(u^2) .
\ee
Thus, we reobtain here, within lattice perturbation theory,
the non-perturbative results discussed before. In the large-$N$ limit
$N\to\infty$, $u\to 0$, with $Nu$ fixed,
$\tilde{g}_\xi^\phi(z)$ is related to $\tilde{g}_\xi(z)$ defined in 
\reff{eq2.38}. Indeed, it is easy to show 
that
\be
\tilde{g}_\xi^\phi(z) =\; A(u)\; \tilde{g}_\xi(B(u) z),
\ee
where
\be
A(u) = \left({6\over Nu\,{\cal C}_{d,1}}\right)^{1/4} + O(u^{3/4}), \qquad\qquad
B(u) = {1\over {\cal C}_{d,0}} \sqrt{6{\cal C}_{d,1}\over Nu} + O(u^{1/2}).
\ee

\subsection{The model without zero mode}

In this case we obtain for $\xi_L^{(2e)}$:
\be
\left(\xi^{(2e)}_L\right)^{-2} = r + \hat{q}^2_{min} + u \Sigma'_L(r) + O(u^2).
\ee
Proceeding as before, we obtain for $\dr\rightarrow 0$, $L\rightarrow\infty$,
$\delta r L^2$ finite
\be
\left(\xi^{(2e)}_L\right)^{-2} = \hat{q}^2_{min} + 
\dr\left[1-\frac{N+2}{6}u\;{\mathcal{C}}_{d,1}\right] + 
\frac{N+2}{6}u\;L^{2-d}{\widehat{\mathcal{I}}}(\dr L^2) + O(u^2)        .
\label{eqxi2d_t}
\ee
Introducing 
\be
\delta\tilde{r}=
\dr + \hat{q}^2_{min} \left[ 1 + \frac{N+2}{6} u\;{\mathcal{C}}_{d,1} \right],
\ee
and expressing $\widehat{\cal I}(\rho)$ in terms of 
$\widetilde{\cal I}(\rho)$, see \reff{widetildeI-def}, we rewrite 
\reff{eqxi2d_t}\ as
\be
\left(\xi^{(2e)}_L\right)^{-2} = 
\delta\tilde{r}\left[1-\frac{N+2}{6}u\;{\mathcal{C}}_{d,1}\right] + 
\frac{N+2}{6}u\;2d\,\left[ \frac{1}{\delta\tilde{r}L^d}+ 
L^{2-d} \widetilde{{\mathcal{I}}}(\delta\tilde{r}L^2)\right] + O(u^2).
\label{eqxi2d_tIII}
\ee
This equation is similar to~\reff{eqxi2a_tII} and therefore, 
$\xi^{(2e)}_L$ has an anomalous scaling behaviour. However, as already noticed 
in the $\sigma$-model, instead of $\delta r$, the scaling function 
contains $\delta\tilde{r}$, indicating the presence of a finite-volume 
temperature renormalization proportional to $L^{-2}$.

\appendix
\section{Some useful functions}
\label{usefulfunc}

We define:
\begin{equation}
\Psi(\lambda_L,L)\equiv\frac{1}{L^d}\sum_{q\in\Lambda_{L}^* -\{0\}}
\frac{1}{\hq^2+\lambda_L}  - \int_{-\pi}^\pi\, {d^dq\over (2 \pi)^d}
  {1\over \hat{q}^2 +   \lambda_L },
\label{defPsi}
\ee
for $\lambda_L\ge 0$.
We wish now to compute the asymptotic behaviour of $\Psi(\lambda_L,L)$ 
for $\lambda_L\to0$, $L\to\infty$, $\lambda_L L^2$ fixed. We begin 
by rewriting $\Psi(\lambda_L,L)$ as 
\be
\Psi(\lambda_L,L) =\, - {1 \over L^d \lambda_L}\,+
\,\sum_{\mathbf{n}\in{\mathbb{Z}}^d-\{0\}} \int_{-\pi}^\pi\,
{d^dq\over (2 \pi)^d}
  {1\over \hat{q}^2 +   \lambda_L } e^{i \mathbf{q}\cdot\mathbf{n} L},
\label{A.2}
\ee
where we have used the Poisson resummation formula
\be
        \sum_{\mathbf{n}\in{\mathbb{Z}}^d} e^{i
          \mathbf{q}\cdot\mathbf{n}L} = \left( \frac{2\pi}{L}\right)^d
        \sum_{\mathbf{m}\in{\mathbb{Z}}^d}\delta^{(d)}\left(\mathbf{q}-
         \frac{2\pi}{L}\mathbf{m}\right)  . 
\ee       
The integral in Eq. \reff{A.2} is ultraviolet convergent because of the
phase factor, and, in the limit we are interested in, the relevant 
region of integration corresponds to $q$ small.
Thus, we can expand in powers of $q$ the denominator and extend the 
integration domain to the whole space. We obtain 
\be
\Psi(\lambda_L,L)\approx L^{2-d} {\cal I}\left(\lambda_L L^2\right),
\label{form3}
\ee
where
\be
{\cal I}(\rho) \equiv 
- {1 \over \rho}\,+
\,\sum_{\mathbf{n}\in{\mathbb{Z}}^d-\{0\}} \int_{\mathbb{R}}\, 
{d^dq\over (2 \pi)^d}
  {1\over q^2 + \rho } e^{i \mathbf{q}\cdot\mathbf{n} }.
\label{defcalI}
\ee
The corrections are of order $L^{-d}$ times a function of $\lambda_L L^2$.
We wish now to compute the behaviour of ${\cal I}(\rho)$ for $\rho \to 0$.
For this purpose, we rewrite
\begin{eqnarray}
{\cal I}(\rho) & = &  - {1 \over \rho}\,+ \int_0^\infty dt\, e^{-t \rho}
 \left[ \sum_{\mathbf{n}\in{\mathbb{Z}}^d -\{0\}}  
\int_{\mathbb{R}} \, {d^dq\over (2 \pi)^d} 
   e^{i\mathbf{q}\cdot\mathbf{n}  - t q^2} \right]  \nonumber \\
 & = &   - {1\over \rho} +  \int_0^\infty { dt \over (4\,\pi\,t)^{d/2}}
\,e^{-\rho t}  \left[ {\cal{B}}^{d}\left(
    \frac{1}{4\,\pi\,t}\right) - 1\right],
\label{last}
\end{eqnarray}
where
\be
{\cal{B}}(s) \equiv \sum_{n=-\infty}^{+\infty}\ e^{-\pi s n^{2}}.
\ee
After the change of variable  $t\mapsto  t/4\pi$,
we have
\begin{eqnarray}
{\cal{I}}(\rho) &=& - {1\over \rho} + {1\over 4 \pi} \left( \int_0^1 +
\int_1^\infty \right)  dt\,
e^{-\frac{\rho}{4 \pi} t} \frac{1}{t^{d/2}} \left[ {\cal{B}}^{d}\left(
    \frac{1}{t}\right) - 1\right]  \nonumber \\ 
&=& - {1\over \rho} + {1\over 4 \pi} \int_1^\infty dt\,
e^{-\frac{\rho}{4 \pi} \frac{1}{t}} t^{d/2 - 
  2} \left[ {\cal{B}}^{d}(t) - 1\right] +\nonumber \\  
& & +\, {1\over 4 \pi} \int_1^\infty dt\, e^{-\frac{\rho}{4 \pi} t}
\frac{1}{t^{d/2}} \left[ {\cal{B}}^{d}\left( \frac{1}{t}\right) - 1\right] ,
\end{eqnarray}
where in the first integral we performed the change of variable $t
\mapsto 1/t$. By using the identity which follows from the Poisson 
resummation formula
\be
{\cal{B}}(s)=\left(\frac{1}{s}\right)^{1/2}\ {\cal{B}}\left(\frac{1}{s}\right),
\label{propB}
\ee
we finally obtain
\begin{eqnarray}
{\cal{I}}(\rho) &=& \frac{e^{-\frac{\rho}{4 \pi}}-1}{\rho} -
\frac{1}{4\pi}\int_1^\infty dt\, t^{-d/2} e^{-\frac{\rho}{4 \pi} t}
+\nonumber\\ 
&+&\frac{1}{4\pi} \int_1^\infty dt\, \left(e^{-\frac{\rho}{4 \pi} t} +
  t^{d/2 -2} e^{-\frac{\rho}{4 \pi} \frac{1}{t}} \right)\left[
  {\cal{B}}^{d}(t) - 1\right]  .\label{A10}
\end{eqnarray}
Note that the first and the third term in ${\cal{I}}(\rho)$\ may
be expanded in integer powers of $\rho$\ around $\rho=0$, while the
second term has an expansion of the form (for $d>0$\ and $d/2$\ not
integer):
\be
\int_1^\infty dt\, t^{-d/2} e^{-\frac{\rho}{4 \pi} t} =
\sum_{k=0}^{+\infty}\, \frac{(-1)^k}{(\frac{d}{2}-1-k)\
  k!}\left(\frac{\rho}{4 \pi}\right)^k \ + \ \left(\frac{\rho}{4\pi}
\right)^{\frac{d-2}{2}}\Gamma\left(\frac{2-d}{2}\right) \; .\label{A11}
\ee  
The generalization for $d/2$ integer is straightforward: 
\begin{eqnarray}
\int_1^\infty dt\, t^{-d/2} e^{-\frac{\rho}{4 \pi} t} & = & \!\! \sum_{k=0,
  \;k\neq\frac{d}{2}-1}^{+\infty}\, \frac{(-1)^k}{(\frac{d}{2}-1-k)\
  k!}\left(\frac{\rho}{4 \pi}\right)^k \nonumber \\
& &  + \
\frac{(-1)^{\frac{d}{2}}}{(\frac{d}{2}-1)!}\left(\frac{\rho}{4\pi}
\right)^{\frac{d-2}{2}}\left[ \gamma + \ln\left(
    \frac{\rho}{4\pi}\right)\right] \; .
\end{eqnarray}
As a consequence of these results, we find that ${\cal I}(\rho)$ 
is finite for $\rho\to 0$, together with its derivatives 
${\cal I}^{(k)}(0)$, $k < \im{\frac{d}{2}}-1$.


We wish now to define a second function
\be
\widehat{\Psi}(\lambda_L,L)  \equiv  \frac{1}{L^d} \,
 \sum_{q\in\Lambda_{L}^* -\{0\}} \frac{1}{\hat{q}^2
 +\lambda_L} - \sum_{n=0}^{\im{\frac{d}{2}}-1}
  (-\lambda_L)^n \int_{-\pi}^\pi\,
  \frac{d^dq}{(2\pi)^d}\! \frac{1}{(\hat{q}^2)^{n+1}},
\label{Psi_N}
\ee
which is defined for $\lambda_L > -\hat{q}^2_{\rm min}$ where 
$q_{\rm min} = (2\pi/L,0,\ldots,0)$, and for $d/2$ not integer. Here
$\im{x}$  denotes the integer part of $x$. 

We are interested, as before, in the asymptotic limit 
$\lambda_L\to0$, $L\to \infty$, $\lambda_L L^2$ finite.
For this purpose we rewrite:
\begin{eqnarray}
\widehat{\Psi}(\lambda_L,L)  &=& 
 \frac{1}{L^d}\sum_{q\in\Lambda_{L}^* -\{0\}} \left\{
   \frac{1}{\hq^2+\lambda_L}-\sum_{n=0}^{\im{\frac{d}{2}}-1}
   \frac{(-\lambda_L)^n}{(\hq^2)^{n+1}}\right\} +\nonumber\\
&=& \sum_{n=0}^{\im{\frac{d}{2}}-1} (-\lambda_L)^n \left\{
  \frac{1}{L^d}\sum_{q\in\Lambda_{L}^* -\{0\}}\frac{1}{(\hq^2)^{n+1}}
  -  \int_{-\pi}^\pi\, \left(
  \frac{dq}{2\pi}\right)^{d} \! \frac{1}{(\hat{q}^2)^{n+1}}\right\}.
\end{eqnarray}
Now, in the first sum we can replace $\hat{q}^2$ by $q^2$ and extend 
the summation over $\mathbb{Z}^d$, since, 
because of the subtracted term, the resulting series
converges. For the second term we immediately see that,
for $1\le n < d/2$,
\be
\frac{1}{L^d}\sum_{q\in\Lambda_{L}^*-\{0\}}\frac{1}{(\hq^2)^n} - 
    \int_{-\pi}^\pi\, \frac{d^dq}{(2\pi)^d} \, \frac{1}{(\hat{q}^2)^n} =\, 
 {(-1)^{n-1}\over (n-1)!} \left.
 {\partial^{n-1}\over \partial\sigma^{n-1}}\,\Psi (\sigma,L)\right|_{\sigma=0}.
\ee
Using the asymptotic expansion of $\Psi(\lambda_L,L)$ we finally obtain
\be
\widehat{\cal I}(\rho) = \lim_{L\to\infty} {1\over L^{2-d}}
\widehat\Psi \left({\rho\over L^2},L\right), \label{IN}
\ee 
where
\be
\widehat{\cal I}(\rho) \equiv  
 \sum_{k=0}^{\im{\frac{d}{2}}-1} {\rho^k\over k!}\, {\cal I}^{(k)}(0) + 
          \sum_{\mathbf{n}\in{\mathbb{Z}}^d-\{0\}} \left\{
            \frac{1}{4\pi^2\mathbf{n}^2+\rho}-\sum_{k=0}^{\im{\frac{d}{2}}-1}
            \frac{(-\rho)^k}{(4\pi^2\mathbf{n}^2)^{k+1}} \right\}.
\label{defL}
\ee
This implies that for $k<\im{d/2}-1$
\be
  \widehat{\cal  I}^{(k)}(0) =  {\cal  I}^{(k)}(0). \label{dk}
\ee 
For positive $\lambda_L$ we wish to evaluate the difference between 
${\cal I}(\rho)$ and $\widehat{\cal I}(\rho)$.
For this purpose we rewrite \reff{last} using \reff{propB}
\be
{\cal{I}}(\rho) = \int_0^\infty { dt \over (4\,\pi\,t)^{d/2}}
\,e^{-\rho t}  \left\{ (4\,\pi\,t)^{d/2} [{\cal{B}}^{d}(4\pi t) - 1] - 1
    \right\}.
\ee
Then, for $d/2$ not integer, since ${\cal I}^{(k)}(0)$ exists for 
$k < {d\over2} - 1$, we rewrite
\begin{eqnarray}
{\cal{I}}(\rho) &= &
  \sum_{k=0}^{\im{\frac{d}{2}}-1} {\rho^k\over k!}\, {\cal I}^{(k)}(0) 
\nonumber \\ 
  && 
+ \int_0^\infty { dt \over (4\,\pi\,t)^{d/2}} 
 \left[e^{-\rho t} - \sum_{k=0}^{\im{\frac{d}{2}}-1} {(- t \rho)^k\over k!}
 \right]
 \left\{ (4\,\pi\,t)^{d/2} [{\cal{B}}^{d}(4\pi t) - 1] - 1
    \right\} 
\nonumber \\
  &= &
  \sum_{k=0}^{\im{\frac{d}{2}}-1} {\rho^k\over k!}\, {\cal I}^{(k)}(0) +
          \sum_{\mathbf{n}\in{\mathbb{Z}}^d-\{0\}} \left\{
            \frac{1}{4\pi^2\mathbf{n}^2+\rho}-\sum_{k=0}^{\im{\frac{d}{2}}-1}
            \frac{(-\rho)^k}{(4\pi^2\mathbf{n}^2)^{k+1}} \right\} 
\nonumber \\
  && 
 - \int_0^\infty { dt \over (4\,\pi\,t)^{d/2}}
 \left[e^{-\rho t} - \sum_{k=0}^{\im{\frac{d}{2}}-1} {(- t \rho)^k\over k!}
 \right] 
\nonumber \\
 &=& \widehat{\cal{I}}(\rho) - {\rho^{d/2-1}\over (4\pi)^{d/2}} 
  \Gamma\left(1 - {d\over2}\right)
\end{eqnarray}
from which, by comparison with \reff{A10} and \reff{A11}, we see that
$\widehat{\cal{I}}(\rho)$ can be extended analytically  around $\rho=0$.


\end{document}